\documentstyle[twoside,fleqn,jhuwkshp,psfig, epsfig]{article}
\newcommand{\etal}{{\it et al.} }

\newcommand{\xte}{XTE J1550$-$564 }
\newcommand{\gtae}{$\buildrel {\lower3pt\hbox{$>$}} \over 
{\lower2pt\hbox{$\sim$}} $}
\newcommand{\ltae}{$\buildrel {\lower3pt\hbox{$<$}} \over
{\lower2pt\hbox{$\sim$}} $}

\begin{document}

\title{Low and High Angular Momentum Accretion Flows in BHCs: Case Study of \xte }
 
\author{Roberto Soria, Kinwah Wu
\address{Mullard Space Science Laboratory,
UCL, Holmbury St.~Mary, Dorking, RH5~6NT, UK.},
Diana Hannikainen \address{Department of Physics and Astronomy,
  University of Southampton, Southampton, SO17~1BJ, UK.},
Mike McCollough \address{SD50, NASA-MSFC, Huntsville, AL 35812, USA.},
Richard Hunstead
\address{School of Physics, University of Sydney, 
  NSW 2006, Australia}}

\begin{abstract}
The 1998 outburst of \xte started with a hard X-ray spike, rising 
in less than a day and declining after 3-4 days; at the same time, 
the soft X-ray flux was rising with a longer timescale ($\sim 10$ days).
We suggest that the soft and the initial hard X-ray emission 
are produced by two different components of the accretion flow: 
a higher angular momentum flow, which forms the disk, and a lower 
angular momentum flow feeding the hot inner region. Thus, we argue that 
the onset of the outburst is determined 
by an increased mass transfer rate from the companion star, 
but the outburst morphology is also determined by the distribution 
of specific angular momentum in the accreting matter.
In \xte, we attribute the initial, impulsive low angular momentum accretion 
to the breaking down of magnetic confinement by the magnetically active 
secondary star. We show that a hard X-ray spike is seen at the onset 
of an outburst in other BHCs. 
\end{abstract}

\maketitle


\section{Introduction}


The BH soft X-ray transient \xte was discovered on MJD~51063 
  (1998 Sep 7) 
  by the ASM on board 
  {\it RXTE} \cite{smi98} 
  and by BATSE  
  on board {\it CGRO} 
  \cite{mcc98}.   
The optical counterpart was identified shortly afterwards
  \cite{oro98}, 
  and a radio source was found at the optical position on MJD~51065 
\cite{cam98},
 with the Molonglo Observatory Synthesis Telescope (MOST), at 843MHz.
From the optical ellipsoidal modulations, 
an orbital period of $1.541\pm0.009$~d was inferred \cite{jai01}.

We used BATSE to monitor the hard X-ray emission (20--100 keV) 
from \xte, from the start of the 1998 Sep outburst until 
{\it CGRO}'s re-entry (2000 Jun). We plot the BATSE and ASM fluxes 
in Fig. 1, together with the ASM hardness ratios. Details of the technique 
used to obtain a daily lightcurve are given in \cite{kwu01}.
We also followed the radio source with MOST for the first 27 days after the 
initial detection. For a preliminary analysis 
of our radio data, see \cite{kwu01,han01,han02}.

\begin{figure*}[t] 
\vspace{-14pt}
\centerline{\psfig{file=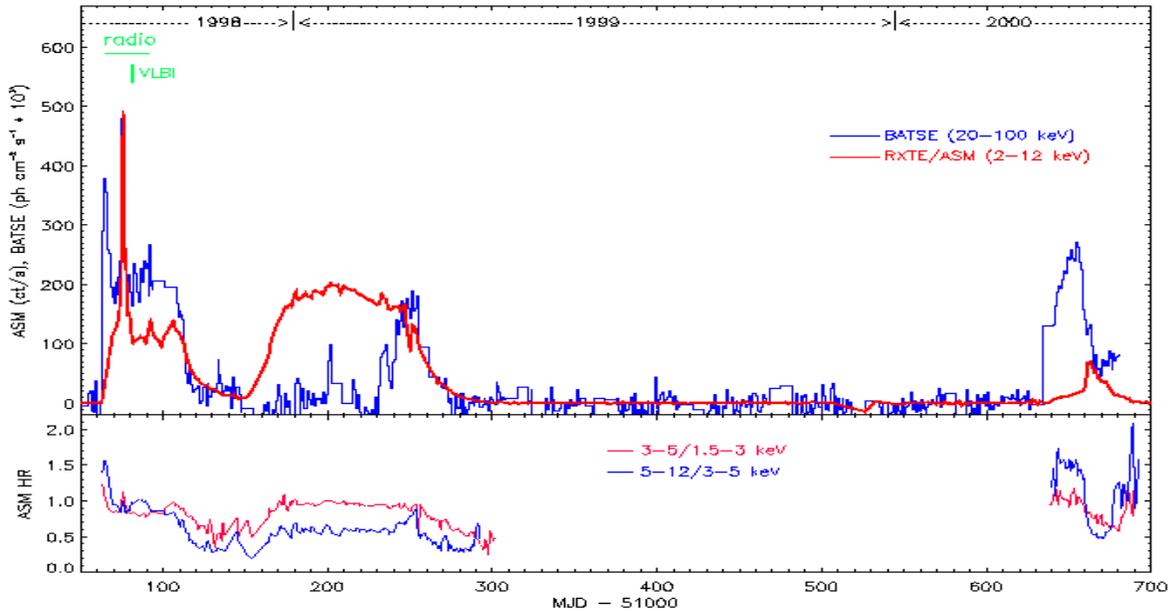,height=7.2in, width=4.0in, angle=270}}
\vspace{-30pt}
\caption{Top panel: BATSE (20--100 keV) and ASM (2--12 keV) fluxes from 1998 Sep 
to 2000 Jun. Bottom panel: ASM hardness ratios for the same period.
}\label{fig:smallfig}
\end{figure*} 

\begin{figure*}[t] 
\vspace{-30pt}
{\epsfig{file=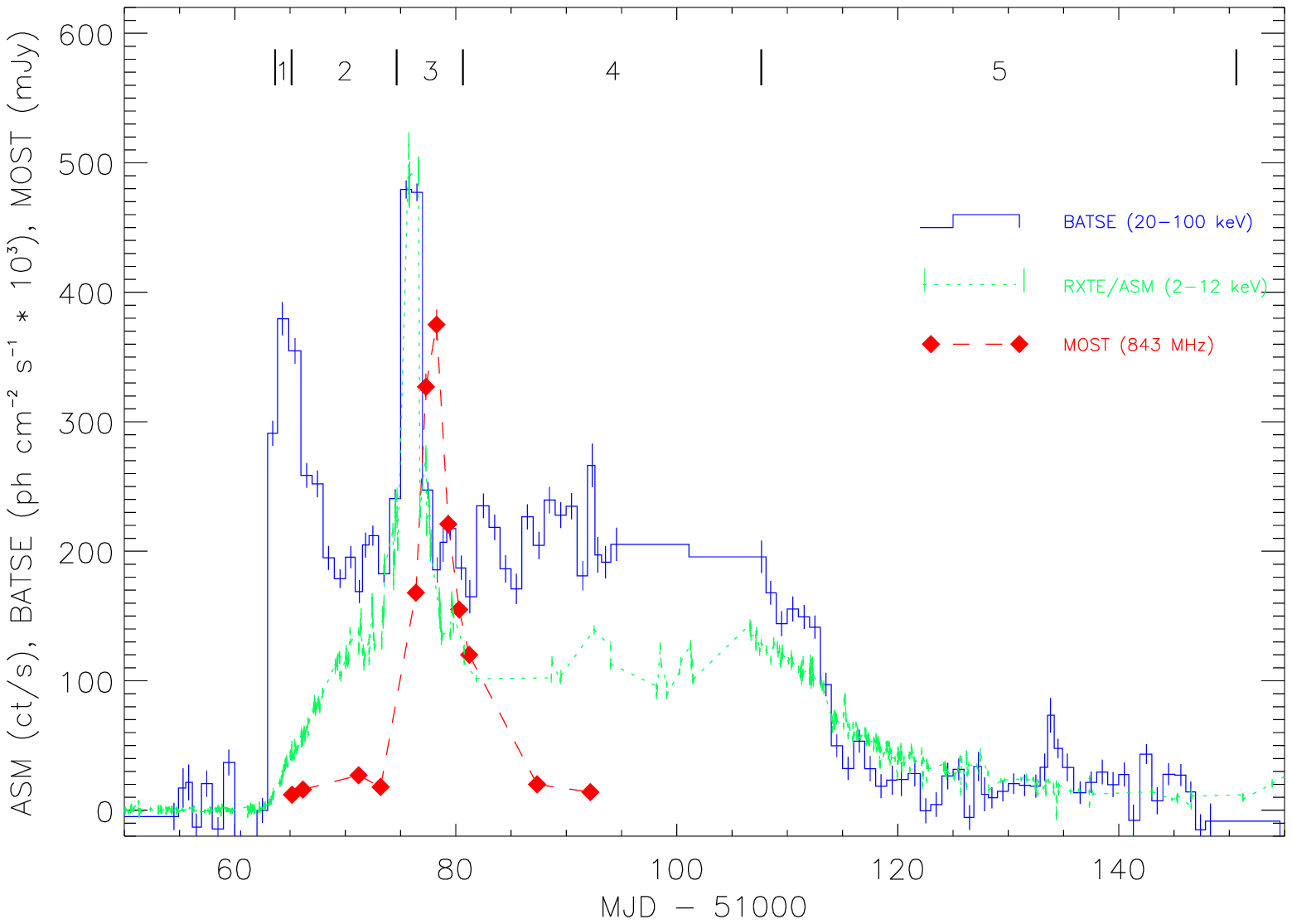,width=6.8in, height=3.92in}}
\vspace{-37pt}
\caption{
Hard X-ray, soft X-ray and radio lightcurves of \xte; see \cite{kwu01} 
for details.
}\label{fig:largefig}
\end{figure*}

\section{Phenomenology of the outburst}  
Remarkable features of the 1998 outburst are: 
\begin{itemize}
\item {\it an initial impulsive rise in the hard X-rays} (Fig.\ 2).
The BATSE (20--100~keV) flux 
  reached about 0.3~ph~cm$^{-2}$s$^{-1}$ within one day, 
  peaking at about 0.4~ph~cm$^{-2}$s$^{-1}$ the next day. 
From {\it RXTE}/PCA$+$HEXTE data \cite{wil01}, 
the 20--200~keV flux peaked at 
$\approx 1.7 \times 10^{37}~{\rm erg}~{\rm s}^{-1}$, for a distance of 
2.5 kpc \cite{san99}.
Assuming an efficiency $\eta \sim 0.1$, 
  the total accreted mass required to account for the initial 
  hard X-ray burst is \ltae $10^{23}$~g;
\item {\it a subsequent giant flare} that occurred almost simultaneously 
      in both hard and soft X-rays 
      about twelve days after the initial rise of the hard X-rays. The radio 
flare occurred at MJD~51077.8$\pm0.1$, $\approx 1.8$~d 
  after the X-ray peak, 
  with the flux density reaching 380~mJy at 843~MHz.
The hard and soft X-ray flux then settled on a plateau, with spectral and timing properties  
similar to the ``very high state'' identified in other BHCs;
\item {\it an exponential decay in the
      soft X-ray flux} after about 40 days. 
      The same decay timescale for the soft flux is seen after 
      the 1998, 1999 and 2000 outbursts.
\end{itemize}


\section{Origin of the hard X-rays}

Hard X-ray emission  
is not thermal emission from an accretion disk (whose maximum 
temperature is $\sim 1$ keV), but is 
produced by inverse Compton scattering of softer photons 
off highly energetic ($E \sim 100$ keV) electrons near the BH.
Different models have been proposed to explain the source 
and location of the hot electrons.

\begin{itemize}
\item Accretion of matter 
with low angular momentum
can produce hard X-rays 
  \cite{igu99,bel00,cha95}.  
The physical justification is that free-falling 
electrons in a quasi-spherical inflow 
can acquire kinetic energies \gtae 100 keV as they approach 
the BH horizon, before reaching their 
circularisation radius. The kinetic energy of the infalling electrons 
powers the Comptonisation process, either directly (bulk-motion 
Comptonisation) or after it has been converted into thermal energy 
(thermal Comptonisation).

\item Alternatively, a hot, optically thin, quasi-spherical flow 
may already exist at small radii before the onset of the outburst 
(ADAF solutions;  \cite{nar95,esi97}). 
The ADAF model may explain residual hard X-ray emission 
seen in some BHCs in quiescence.

\item Other physical mechanisms for the production and storing of  
the Comptonising electrons (eg, a magnetically heated corona 
on top of the disk) may be present 
at later stages of the outburst, and may explain 
the hard X-ray emission in the plateau phase (MJD 51080 -- 51100). 
But it is unlikely that they could operate at the onset 
of the outburst, before a disk was fully formed.

\end{itemize}

\section{Low angular momentum inflow}

We suggest that the initial hard X-ray spike 
is due to impulsive accretion of matter with low specific angular 
momentum, over the first few days. Meanwhile, high angular 
momentum matter would also be accreted through normal Roche lobe (RL) overflow.
{\it The outburst behaviour is determined by the mass transfer rate} 
  (which sets the energetics of the burst) 
{\it and the angular momentum distribution of the accretion flow}
  (which sets the initial spectral properties and the X-ray rise time).

In the ADAF scenario, the optically thin inner flow would be 
already present when the outburst starts, and 
would be responsible for the initial hard emission. The 
subsequent spectral softening  
  is attributed to the collapse of the ADAF 
  into an optically thick disk 
  when the mass accretion rate increases. 
Only one parameter determines the morphology of the outburst 
in the ADAF model: the mass accretion rate. 

\section{Magnetic confinement}

Wind accretion is a typical example of low angular momentum 
accretion flows. In fact, the X-ray spectrum of \xte during the initial 
hard phase is very similar to the spectrum of Cyg X-1 in its 
low/hard state, when wind accretion is thought to dominate over 
RL overflow \cite{wil01}. However, 
the companion star in \xte is likely to be a K subgiant, 
therefore it cannot produce a wind. Is there an alternative source 
for the subkeplerian accretion flow?

We notice that fast-rotating K subgiant stars in binary systems 
tend to have a strong magnetic field. 
The main physical reason is   
that, while the tidally-deformed stellar envelope 
  can be locked in synchronous rotation with the orbit, 
  the degenerate stellar core 
  may not attain perfect rotational synchronism.
The differential rotation leads to dynamo action and magnetic activity.

For example, RS CVn systems 
are magnetic close binaries containing a G/K subgiant 
with a multi-kilogauss global dipolar field \cite{don92,vog99}. 
Given that the orbital period in \xte is even shorter, 
and tidal deformation of the companion is even stronger than in 
typical RS CVn systems, 
we suggest that the K subgiant companion in \xte is also strongly magnetic. 

Extended regions of cool, optically-thin, magnetically confined material 
  around a K star are a common feature of RS CVn systems  \cite{hal94}. 
We suggest that the same magnetic confinement occurs in \xte.
In fact, the mass of the confined gas 
   can be even higher here, 
   because the magnetic companion star is close to 
   filling its RL. 
We show in \cite{kwu01} that a mass $> 10^{23}$~g 
can be stored in this ``magnetic bag'', enough to account for 
the initial hard X-ray burst.
By analogy with RS CVn systems, 
  the confined gas in \xte 
  may be located well beyond the L$_1$ point, 
  deep into the RL of the primary, 
  but corotating with the binary (therefore with lower 
angular momentum than the material accreted via RL overflow).



The start of the outburst 
is determined by an increased mass loss from the secondary star. 
Part of this matter goes through L$_1$, is accreted with a higher 
specific angular momentum 
and is responsible for the formation of the disk. Another component is 
temporarily channelled 
into the magnetic bags above the companion star, until the confinement 
breaks down; it then 
falls towards the accreting primary with lower angular momentum, 
feeding the hot, hard X-ray emitting region.

\section{A subclass of BHCs with a hard spike}

\xte is not the only system showing a hard X-ray spike 
  at the onset of an outburst.   
XTE J1859$+$226 \cite{woo99,mcc99}
  and XTE J2012$+$381 \cite{vas00} are two other good examples.  
In these three systems, 
  the hard X-ray flux reached a maximum within the first day after detection, 
  and then declined after about 4 days.
The soft X-ray flux increased more steadily and at a slower pace, 
  for $\sim 10$ days. 
The similarity in their X-ray spectral properties 
  can be seen from their hardness-ratio plots (Fig. 3).
There is probably a subclass of BH transients,  
  sharing a similar physical mechanism 
  that gives rise to a strong hard X-ray spike  
  at the onset of an outburst.  
A more detailed analysis of the hard X-ray properties 
  of this type of systems will be presented elsewhere.

\section{Conclusions}

We discuss evidence of a distribution of angular momenta in the accretion 
flow onto the BH transient \xte. We argue that the initial hard X-ray burst 
is due to a subkeplerian component, while matter accreted through the Roche  
lobe is responsible for the formation of a disk. 
We propose that the companion star 
is magnetically active, and its magnetic field creates a ``bag'' 
capable of confining $\sim 10^{23}$ g of gas inside 
the Roche lobe of the primary, corotating with the binary. 
When the confinement breaks down, this low angular 
momentum gas reaches the inner region near the BH horizon 
on a free-fall timescale.
We suggest that the onset of the outburst is determined 
by an increased mass transfer rate from the companion star, 
but the outburst morphology is determined by the distribution 
of specific angular momentum in the accreting matter.

\begin{figure}[t] 
\vspace{10pt}
\centerline{\epsfig{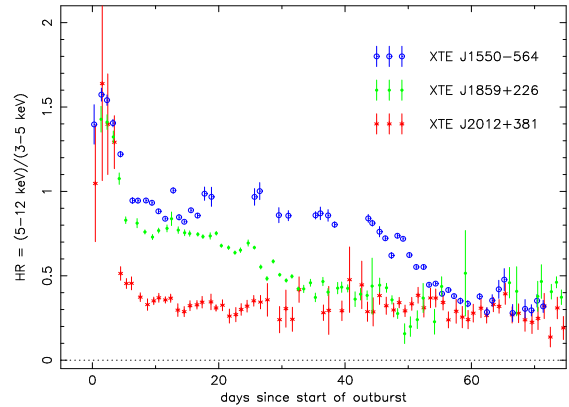}}
\vspace{-20pt}
\caption{
Evolution of 
the {\it RXTE}/ASM hardness ratios (5--12 / 3--5 keV band) 
during recent outbursts of the BHCs \xte (open circles), 
XTE J1859$+$226 (filled circles) and XTE J2012$+$381 (asterisks). 
See  \cite{kwu01} for details.
}\label{fig:smallfig}
\end{figure}



\normalsize

 
\end{document}